# Identifying reasoning patterns in games

**Dimitrios Antos**
School of Engineering and Applied Sciences
Harvard University
Cambridge, MA 02138
dantos@eecs.harvard.edu

**Avi Pfeffer**
School of Engineering and Applied Sciences
Harvard University
Cambridge, MA 02138
avi@eecs.harvard.edu

## Abstract

We present an algorithm that identifies the reasoning patterns of agents in a game, by iteratively examining the graph structure of its Multi-Agent Influence Diagram (MAID) representation. If the decision of an agent participates in no reasoning patterns, then we can effectively ignore that decision for the purpose of calculating a Nash equilibrium for the game. In some cases, this can lead to exponential time savings in the process of equilibrium calculation. Moreover, our algorithm can be used to enumerate the reasoning patterns in a game, which can be useful for constructing more effective computerized agents interacting with humans.

## 1 INTRODUCTION

Games are strategic interactions between agents that have different capabilities, information and objectives. Such games, which often involve uncertainty about the world, have become vitally important in computer science as the basis for describing multiagent interaction. The standard solution concept for games is Nash equilibrium. Solving a game is usually taken to mean finding a Nash equilibrium which will specify the strategies of agents. Unfortunately, however, computing Nash equilibrium is a serious bottleneck to using the game theoretic approach; as Daskalakis et al. [2006] have shown, calculating a single Nash equilibrium is PPAD complete. The UAI community has developed graphical representations such as graphical games [Kearns et al. 2001], action graph games [Jiang and Leyton-Brown 2006] and multi-agent influence diagrams (MAIDs) [Koller and Milch 2001], all of which have structural properties that assist in the solution of games. Even in these frameworks, however, computing Nash equilibria can still be very hard.

This paper addresses this bottleneck in two ways. First, it presents a method for simplifying a game in order to make it easier to solve. The method works by analyzing a MAID to discover the reasoning patterns that apply in the given situation. A reasoning pattern is a form of argument that can lead to or motivate a decision. Pfeffer and Gal [2007] showed that all reasoning patterns in MAIDs fall into one of four graphical categories. Furthermore, they showed that if no reasoning pattern holds for a particular decision, the agent making the decision has no reason to prefer one action over another.

We present an algorithm that identifies whether reasoning patterns hold for different decisions, and simplifies the game if they do not. Our algorithm relies on the definitions of reasoning patterns in [Pfeffer and Gal 2007] and also integrates insights of Koller and Milch [2008] that identify cases in which edges can safely be removed from a MAID. We use an iterative procedure in which the MAID is repeatedly simplified. We prove that our algorithm is correct and that it always results in a maximally simplified MAID. In particular, the order in which nodes and edges are considered for removal does not matter. We show that our algorithm has polynomial running time, and also demonstrate how memoization improves the complexity of the algorithm. We present an example showing that our algorithm leads to significant savings in the cost of solving a MAID, in some cases exponential savings.

The second way we address the bottleneck is by providing support for non-equilibrium decision making. Here the goal is to develop good strategies for games which are not necessarily Nash equilibria. We extend our algorithm to identify all the reasoning patterns in a game. This might be helpful in a number of ways. First, the reasoning patterns could be presented to a human decision maker who makes the ultimate decisions. By examining the reasoning patterns, the decision maker can weigh the different arguments for making decisions and come up with a good decision, based on his or her beliefs about how others are making their decisions. Second, reasoning patterns might provide the basis for automatic approximate solution of games. Instead of solving a single large game, one might solve a number of simpler games and combine the solutions to produce strategies for the large game. Third, reasoning patterns can help characterize how people actually play in games, and thereby help in developing strategies that best respond to people's play. For example, the reasoning pattern known as signaling brings about issues of trust, while the pattern known as manipulation focuses agents on the reciprocal aspects of their interaction. By making these social aspects of reasoning explicit, reasoning patterns facilitate understanding and modeling behavior. Finally, even in games in which an equilibrium can be computed, reasoning patterns may be useful in explaining the equilibrium to human decision makers. Instead of a dry suggestion "choose decision rule $\delta$ because it is a Nash equilibrium," the computer decision support system will be able to say "choose $\delta$ because it will encourage another agent to help you without sacrificing too much of your own utility."

## 2  PRELIMINARIES

Multi-Agent Influence Diagrams (MAIDs) are an extension of Influence Diagrams to the multi-agent case [Koller and Milch 2001]. A MAID is a directed acyclic graph with three types of nodes: *Chance* nodes, represented as circles, contain conditional probability distributions over their domains given their inputs. *Decision* nodes, represented as rectangles, are used to denote choices belonging to particular agents. Edges incoming to a decision node reveal the information that is available to the agent at the time of his decision. Finally, *utility* nodes, which are diamond-shaped and also belong to particular agents, represent deterministic functions from the values of their parents to real numbers. A *strategy* $\sigma_A$ for an agent $A$ in a MAID is a set of decision rules, one for each decision node belonging to $A$, which maps each configuration of its parents $\mathbf{Pa}(D)$ to a probability distribution over the actions in its domain $\mathrm{Dom}(D)$. Similarly, a *strategy profile* $\sigma$ for a MAID is a set containing one strategy for each decision in the MAID. A strategy profile $\sigma$ defines a probability distribution over all the MAID's nodes $P^\sigma(\mathbf{C}, \mathbf{D}, \mathbf{U}) = \prod_{c_i} Pr(c_i|\mathbf{Pa}(c_i)) \prod_{d_j} \sigma_j(d_j|\mathbf{Pa}(d_j)) \prod_{u_k} \mathrm{I}[u_k = U_k(\mathbf{Pa}(u_k))]$, where $\mathbf{C}$, $\mathbf{D}$ and $\mathbf{U}$ are the chance, decision and utility nodes in the MAID and $\mathrm{I}[x]$ is the indicator of $x$. The *expected utility* of an agent $A$ under strategy profile $\sigma$ is given by $\mathbf{EU}^\sigma(A) = \sum_{U \in U_a} \sum_u Pr^\sigma(U = u)u$. Solving a MAID requires calculating a Nash equilibrium, i.e., a strategy profile such that no agent may hope to increase his expected utility by unilaterally deviating from his chosen strategy.

In their paper, Koller and Milch [2008] describe an algorithm for simplifying the structure of a MAID by removing edges that are incoming to decision nodes. In particular, their method relies on the fact that, whenever a parent of a decision node is d-separated from its utilities given the decision and its other parents, then one can remove the edge connecting this parent to the decision. Our approach for simplifying a game is based on examining the *reasoning patterns* of agents. These are surprisingly simple graphical properties that all decision nodes are expected to possess in a MAID, if the strategies of these nodes are important to its solution. We now define the concepts of a motivated and effective decision node:

**Definition 1.** *A decision node D of agent A is called* motivated *if, for some configuration q of its parents,* $\mathbf{EU}^{<\sigma_{-A}, d_1>}(A|q) \neq \mathbf{EU}^{<\sigma_{-A}, d_2>}(A|q)$ *for* $d_1, d_2 \in \mathrm{Dom}(D)$.

Intuitively, a motivated node is one where the agent cares about his decision. A non-motivated node, by contrast, is one where the agent has no reason to choose one action over another, no matter what the information available to him might be.

**Definition 2.** *A decision node d of agent a is called* effective *if it participates in at least one reasoning pattern.*

We briefly and informally present the four reasoning patterns below. The reader is referred to [Pfeffer and Gal 2007] for formal definitions.

- Direct effect: This reasoning pattern is present when the agent can directly or indirectly (but not through another agent's decision) influence his own utility. An example of direct effect can be seen in Figure 1a.

- Manipulation: Here an agent $A$ may exert influence on another agent $B$, whose utility $A$ can affect. Agent $B$ is "manipulated" to do what $A$ wants her to, through his effect on her utility. See Figure 1b for an example of manipulation.

- Signaling: Here $A$ has access to information that is valuable to $B$, i.e., affects her utility, and may choose an action so as to communicate this information to $B$. The reason why he might do that is because $B$, upon seeing his signal, will be expected to choose an action that is favorable to him. An example is shown in Figure 1c.

- Revealing-denying: In this reasoning pattern the agent $A$ has the ability to allow or obstruct the

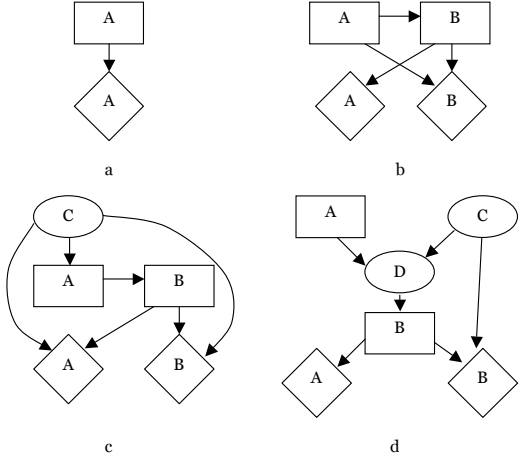

Figure 1: Examples of reasoning patterns

flow of information to another agent B. This way, he effectively increases or reduces the level of certainty B has when making her decision, and thus he may elicit a more favorable response to him. An example is shown in Figure 1d.

To obtain their results, Pfeffer and Gal restrict the strategy space by considering only *well-distinguishing* (WD) *strategies*. This set contains all strategies that do not differentiate between actions based on information that is irrelevant to the decision. Pfeffer and Gal show that WD strategies always contains a Nash equilibrium. Moreover, under WD strategies, the following theorem holds:

**Theorem 1.** *Assuming all agents play WD strategies, if a decision node is motivated, then it is always effective* (proven in [Pfeffer and Gal 2007]).

## 3 SIMPLIFYING MAIDs

The discussion of reasoning patterns suggests that, under some circumstances, some decision nodes will be unmotivated. In this section we present an algorithm for identifying unmotivated nodes; this will allow us to simplify a MAID. We will be able to turn these nodes into chance nodes and remove all edges incoming to them.

The algorithm, presented as Algorithm 1, receives as input a MAID $M$ and returns a simplified version of it. It works by iteratively going through two phases: The first (lines 6-17) is a reasoning pattern identification phase, where for each decision node the algorithm tries to establish that it is *motivated*, i.e., at least one reasoning pattern holds for it. Non-motivated decision nodes are those in which any action taken has no effect on the decision maker's payoff. These are replaced by chance nodes with uniform conditional probability distributions and any edges incoming to them are removed. The second phase (lines 18-20) is an edge pruning phase, which uses the algorithm of [Koller and Milch 2008].

---
**Algorithm 1** The simplification algorithm
---
**Input:** MAID $G$
1: $D \leftarrow$ decision nodes in $G$
2: **for** $d$ in $D$ **do**
3:   $effective(d) \leftarrow true$
4: **repeat**
5:   $retracted \leftarrow false; simplified \leftarrow false$
6:   {identification phase}
7:   **repeat**
8:     $changed \leftarrow false$
9:     **for** $d$ in $D$ **do**
10:       **if** not (**df**($d$) or **man**($d$) or **sig**($d$) or **rev**($d$)) **then**
11:         $effective(d) \leftarrow false$
12:         $simplified \leftarrow true$
13:         $changed \leftarrow true$
14:         remove edges incoming to $d$
15:         make $d$ a chance node w/uniform distr.
16:         discard memoization database
17:   **until** $changed = false$
18:   {pruning phase}
19:   **if** **retract_edges**($G$) **then**
20:     $retracted \leftarrow true$
21: **until** $retracted = simplified = false$

---

The above algorithm uses procedures **df**, **man**, **sig**, **rev** and **retract_edges**, which are defined below. The first four of these correspond to detecting the existence of the respective four reasoning patterns, whereas the fifth implements the algorithm in [Koller and Milch 2008]. The algorithm terminates when a full iteration (identification and pruning phases) causes the graph to remain unchanged.

All these procedures are built upon simple graph reachability or d-separation operations, all of which can be efficiently computed in polynomial time. For example, the procedure **directedDecisionFreePath**$(x, y)$ is simply implemented by a breadth- or depth-first search. Certain more "sophisticated" procedures are used to search for a path that satisfies particular properties. For example, **effectivePath**$(x, y, W)$ is used to search for a path from $x$ to $y$ on which all decision nodes $d$ have $effective(x) = true$ and, moreover, the path is not blocked by the set of nodes $W$. Here *blocking* is interpreted as in any Bayesian network.

**df**($d$)
1: $U \leftarrow$ utility nodes belonging to the owner of $d$
2: **for** $u$ in $U$ **do**

3:  **if directedDecisionFreePath**(u, D) **then**
4:      **return** true
5: **return** false

**man**(d)
1: $U \leftarrow$ utility nodes belonging to the owner of $d$
2: $N \leftarrow$ decision nodes reachable by $d$ through a directed decision-free path
3: **for** $u$ in $U$ and $n$ in $N$ **do**
4:   $U' \leftarrow$ utilities belonging to the owner of $n$
5:   **for** $u'$ in $U'$ **do**
6:     **if directedEffectivePath**(n, u) and **directedEffectivePathNotThrough**(d, u', n) **then**
7:       **return** true
8: **return** false

**sig**(d)
1: $U \leftarrow$ utility nodes belonging to the owner of $d$
2: $N \leftarrow$ decision nodes reachable by $d$ through a directed decision-free path
3: **for** $u$ in $U$ and $n$ in $N$ **do**
4:   $U' \leftarrow$ utilities belonging to the owner of $n$
5:   $w' \leftarrow$ all parents of $n$ that are not descendants of $d$
6:   **for** $u'$ in $U'$ **do**
7:     **if directedEffectivePath**(n, u) **then**
8:       $A \leftarrow$ ancestors of $d$
9:       **for** $a$ in $A$ **do**
10:        **if backDoorPath**(a, u', w') **then**
11:          $w \leftarrow$ all parents of $d$ that are not descendants of $a$
12:          **if effectivePath**(a, u, w) **then**
13:            **return** true
14: **return** false

**rev**(d)
1: $U \leftarrow$ utility nodes belonging to the owner $d$
2: $N \leftarrow$ decision nodes reachable by $d$ through a directed decision-free path
3: **for** $u$ in $U$ and $n$ in $N$ **do**
4:   $U' \leftarrow$ utilities belonging to the owner of $n$
5:   $w \leftarrow$ all parents of $n$ that are not descendants of $d$
6:   **for** $u'$ in $U'$ **do**
7:     **if directedEffectivePath**(n, u) and **frontDoorIndirectPath**(d, u', w) **then**
8:       **return** true
9: **return** false

**retract_edges**(d)
1: $InfEdges \leftarrow$ all edges incoming to decision nodes in $G$
2: $removed \leftarrow false$
3: **for** $(x, y)$ in $InfEdges$ **do**
4:   $disabled((x, y)) \leftarrow true$
5: $D \leftarrow$ all decision nodes in $G$
6: **repeat**
7:   $change \leftarrow false$
8:   **for** $d$ in $D$ **do**
9:     $Parents(d) \leftarrow$ parents of $d$
10:    $Utilities(d) \leftarrow$ utilities of $d$
11:    **for** $p$ in $Parents(d)$ and $u$ in $Utilities(d)$ **do**
12:      $w(p, d) \leftarrow$ all parents of $d$ except for $p$
13:      **if** not **dSeparUseEnabled**(p, u, w(p, d)) **then**
14:        $disabled((p, d)) \leftarrow false$
15:        $change \leftarrow true; removed \leftarrow true$
16: **until** $change = false$
17: remove all disabled edges from $G$
18: **return** $removed$

**directedDecisionFreePath**($x_1, x_2$)
 return true if there is a directed, decision-free path from $x_1$ to $x_2$

**directedEffectivePath**($x_1, x_2$)
 return true if there is a directed path from $x_1$ to $x_2$ in which all decision nodes, except perhaps the first node of the path, are effective

**effectivePath**($x_1, x_2$)
 return true if there is an undirected path from $x_1$ to $x_2$ in which all decision nodes, except perhaps the first node of the path, are effective

**directedEffectivePathNotThrough**($x_1, x_2, Y$)
 return true if there is a directed effective path from $x_1$ to $x_2$ that does not go through any of the nodes in $Y$

**backDoorPath**($x_1, x_2, W$)
 return true if there is a back-door path from $x_1$ to $x_2$ that is not blocked by $W$

**frontDoorIndirectPath**($x_1, x_2, W$)
 return true if there is a non-directed front-door path with converging arrows at some node from $x_1$ to $x_2$ that is not blocked by $W$

**dSepartUseEnabled**($x_1, x_2, W$)
 return true if $x_1$ is d-separated from $x_2$ given $W$, by using only edges $e$ having $disabled(e) = false$

A back door path in the above methods is defined as an undirected effective path where the first edge comes into the first node. A front door indirect path is an undirected effective path where the first edge comes out of the first node and, moreover, the path has converging arrows at some node.

### 3.1 Proof of correctness

We wish to show that our algorithm performs a legitimate simplification $M'$ of the input MAID $M$, given our assumptions of how agents reason about available information. We also care about the effect of the order under which nodes are being eliminated, in order to guarantee that the maximum possible number of non-effective nodes are detected and removed.

**Definition 3.** *A simplification $M'$ of a MAID $M$ is* legitimate *if all the Nash equilibria of $M'$ are also Nash equilibria of $M$, in the sense that all nodes in $M$ that have not been eliminated do not have an incentive to deviate from their equilibrium strategy in $M'$ and all nodes that are not effective play in $M$ according to a fully-mixed, uniform strategy.*

We begin by first proving that, if the algorithm marks effective and non-effective nodes correctly, all Nash equilibria of the simplified MAID $M'$ are also Nash equilibria of the original MAID. More precisely, since the original game also contains the decision nodes that were eliminated (and replaced by chance nodes) we need to show that extending the equilibria of $M'$ by adding fully mixed uniform strategies for all non-motivated decisions yields a Nash equilibrium of $M$.

**Theorem 2.** *Let $D$ be the decision nodes of the original MAID $M$ and $D'$ be the corresponding decision nodes in the simplified MAID $M'$. If $\sigma'$ is a Nash equilibrium of $M'$ then construct $\sigma$ by adding fully mixed uniform strategies for all decision nodes in $D - D'$. Then $\sigma$ is a Nash equilibrium of $M$.*

*Proof.* We prove this by contradiction. Let there be an agent with a decision node $a$ who wishes to deviate from $\sigma_a$; there are two cases: either $a \in D'$ or $a \in D - D'$. In the first case, if the agent owning $a$ wants to deviate to a strategy $\sigma_a^1$ in $M$ then she would deviate to $\sigma_a^1$ in $M'$ as well, contradicting our assumption that $\sigma'$ is a Nash equilibrium in $M'$. In the second case, $a$ was marked as non-effective, therefore $a$ is not motivated, by Theorem 1. By the definition of a non-motivated node, $\mathbf{EU}^{<\sigma_{-a},d_1>}(a,\mathbf{q}) = \mathbf{EU}^{<\sigma_{-a},d_2>}(a,\mathbf{q})$ for every pair of actions $d_1$, $d_2$ of $a$ and every configuration $\mathbf{q}$ of its informational parents. Therefore $a$ provides with a fully mixed uniform strategy the same payoff as from any possible deviation from it. Therefore $\sigma$ is a Nash equilibrium of $M$. The above reasoning holds even in cases where the agent of $a$ owns other decision nodes besides $a$, from which he might try to simultaneously deviate. The reason is that $a$ is non-motivated due to strictly graphical properties of the MAID, not due to any particular parameters employed in other chance or decision nodes; thus strategies followed by any other agent—including the owner of $a$—elsewhere cannot cause it to become motivated. □

We then prove that our algorithm eliminates nodes correctly and maximally, irrespective of order. We do this first by looking at the four procedures that detect reasoning patterns. These work by explicitly following the definitions of the four reasoning patterns, so they are correct (we omit the details). We then look at the first phase of the algorithm (lines 6-18). First, however, we need the following lemma.

**Lemma 1.** *If $R_n^E$ is the set of reasoning patterns that hold for a decision node $n$ when the set of edges in the MAID is $E$, then $R_n^{E'} \subseteq R_n^E$ for all $E' \subseteq E$.*

*Proof.* Consider a MAID with edges $E$ and a set $R_n^E$ of reasoning patterns holding for decision node $n$. Now remove an edge $e \in E$, such that the MAID now has edges $E' = E - \{e\}$. Suppose now, for the sake of contradiction, that $R_n^{E'} \nsubseteq R_n^E$. This means that a reasoning pattern $r$ did not exist under $E$ but exists under $E'$. Take the paths $P_r$ of this reasoning pattern and let $E(P_r)$ be the set of their edges. Clearly, $E(P_r) \subseteq E' \subset E$ so all the paths the reasoning pattern $r$ depends on existed in the original MAID with edges $E$. Thus the only reason why $r$ did not hold under $E$ was that one or more of its paths were blocked at some node. Let $p \in P_r$ be one such path with blocking set $W_p$ ($W_p = \emptyset$ if the definition of the reasoning pattern required no d-separation properties to hold for that path) and let $b$ be the node where $p$ was blocked by $W_p$. If $p$ has non-converging arrows in $b$ then $b \in W_p$, so $p$ should be blocked again under $E'$, since no nodes were removed, only an edge. If $p$ has converging arrows in $b$ then it means that neither $b$ nor any of its descendants were in $W_p$. But removing $e$ can neither add $b$ to $W_p$, nor cause the set of its descendants to grow. Therefore, under $E'$, too, $W_p$ will block $p$ at $b$ and therefore our argument is contradictory. □

**Lemma 2.** *If a node is identified in some identification phase under some order, it will be so identified under any order.*

*Proof.* Let $n_1,...,n_k$ be an order of identifying non-effective nodes and $n'_1,...,n'_k$ be a different order. Also define as $E_n$ the set of edges in the MAID after node $n$ has been eliminated and as $E$ the edges in the MAID in the beginning, before any elimination takes place. Suppose that under the new order node $n_1$ is placed at position $h$. Then, by Lemma 1, in the identification phase and under the new order $n_1$ will be eliminated irrespective of $h$, since $E'_h \subseteq E$ and node $n_1$ was eliminated under $E$. We then reason by induction. Now assume that $n_1,...,n_i$ have been eliminated. Then $n_{i+1}$ will be eliminated at the latest in the next phase after all of $n_1,...,n_i$ have been eliminated, because the set of edges present will be a subset of $E_i$. □

For the second phase of the algorithm (pruning), this is exactly implemented as in [Koller and Milch 2008], which contains the proof of its correctness.

**Theorem 3.** *The algorithm produces a correct and maximal simplification of a MAID.*

*Proof.* We know that the operation of each phase is correct. Moreover, we know the pruning phase only removes edges. Thus, by Lemma 1, it does not matter in the context of the identification phase on which iteration of the algorithm the pruning phase removes an edge $e$, as long as it eventually removes it (on some iteration). Thus the only thing we need to establish is that no operation in the identification phase might ever prevent an edge from being removed in the pruning phase.

The pruning phase works by testing for certain d-separation properties, while the identification phase only removes edges (never adds). Thus, in spirit similar to the proof of Lemma 1, if the MAID has edges $E$ in the beginning of the pruning phase and an edge $e = (x, y)$ is removed during its execution, then if the MAID had edges $E' \subset E$ then $e$ would still be removed. If $e$ was removed with edges $E$ then $x$ was d-separated from the utility nodes of $y$ given $\{y\} \cup \{d : (d, y) \in E, d \neq x\}$, by definition of [Koller and Milch 2008]'s algorithm. Now under the smaller set of edges $E'$ it is the case that $x$ must, again, be d-separated from $y$'s utility nodes, since the removal of any edge in $E - E'$ cannot have made these two less separated. Therefore, no matter in which order we execute the two phases and no matter what their intermediate results are, the end product is the same.

Furthermore, the iteration of identification and pruning phases will terminate. Neither adds an edge and there are at most $E$ edges to remove, so the process will eventually terminate. □

We have shown that our algorithm's operations are consistent with our assumptions and that it will always return a maximally simplified MAID. In the following section we analyze its complexity.

### 3.2 Algorithm Complexity

The complexity of the algorithm is easy to estimate. We first begin with the bottom-level procedures which are path operations. Those that do not involve a non-empty blocking set are instances of graph reachability, which can be performed in $O(E+N) = O(E)$ time, assuming $E > N$. If the blocking set is non-empty, however, every time a path is expanded one needs to check that it is not blocked at that node. This can be done using an algorithm such as BayesBall [Shachter 1998], which is $O(E)$.

We also use *memoization* to improve the computation of blocking properties along the paths. In particular, whenever we query whether a path with converging arrows at at a node $b$ is blocked by a set $W$, we store the result $blocked(b, W)$ in a hash table. Subsequent queries for the same node and blocking set first check the hash table for an already computed result and only execute the full operation (costing $O(E)$) if needed. After each iteration, since the structure of the graph has changed, we drop all memoized entries (line 16).

**Theorem 4.** *The algorithm simplifies the MAID in time $O(D^2N^2E)$, where $D$ is the number of decision nodes in the graph.*

*Proof.* We have established that all path operations take polynomial time. In particular, suppose $C$ and $U$ are the number of chance and utility nodes. Then procedure **df** performs $O(U)$ simple path operations, so it costs $O(UE)$ in the worst case. Manipulation (**man**) performs $O(DU^2)$ simple operations, for a total cost of $O(DEU^2)$. Signaling (**sig**) requires certain paths to satisfy blocking properties and performs $O(CDU^2)$ of those, for a total worst-case cost of $O(CDU^2E^2)$. Here the $E^2$ results from the following: For every combination of nodes related to the signaling patterns we need to find certain paths with graph reachability ($O(E)$); for each such path, at every step we need to check for blocking properties, which adds another $O(E)$, for at total of $O(E^2)$. Finally, revealing-denying (**rev**) performs $O(DU^2)$ blocking-sensitive path operations and thus has worst case cost of $O(DU^2E^2)$.

We see that signaling is the most expensive of these operations. However, with memoization, its worst-case complexity can be reduced. We reason as follows. There can be a total of $O(DN)$ blocking sets required for the purposes of identifying reasoning patterns in the graph (for every decision node there can be one blocking set including all of its parents but one, and there are $O(N)$ parents per decision). Thus, even if we were to calculate blocking properties for all nodes and possible sets, the time per iteration would be bounded by $O(DN^2E)$. In a similar fashion, the time complexity for revealing-denying identification can be bounded.

The algorithm as a whole also performs at most $O(D)$ iterations in the outer loop. This is because if two consecutive iterations eliminate no nodes the algorithm by definition terminates, since the pruning phase of the second iteration will remove no edges. Therefore the total cost of the algorithm is polynomial and on the order of $O(D^2N^2E)$. □

Of course it has to be noted that the expected performance of our algorithm is likely to be much better than its worst-case bound calculated above. In particular, the evaluation of the *if* structure in line 10 is short-circuited, meaning the expensive **sig** operation is only evaluated if **df** and **man** are false, since it is sufficient to show that a node participates in one reasoning pattern to be effective. Moreover, real games very likely have much fewer reasoning patterns mainly consisting of direct effects and manipulations, especially after the pruning phases of the first iterations have reduced the number of edges in the graph that are important to signaling and revealing-denying patterns.

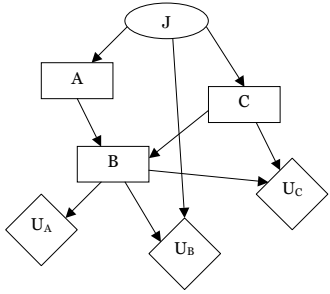

Figure 2: MAID for our simple example

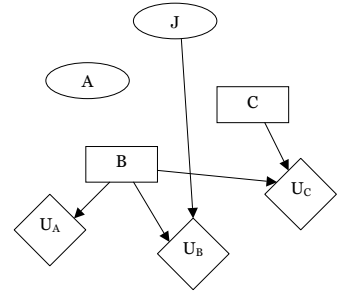

Figure 3: Simplified example MAID

## 4 AN EXAMPLE

Consider the following game. Player $A$ is handed a card, which can be either High ($H$), Medium ($M$) or Low ($L$). Another player ($B$) must guess what type of card $A$ was given; upon a correct guess she wins $10, otherwise she gets nothing. $A$ can tell her what the value of the card is, not necessarily truthfully. $A$'s payoff is $10, $5 or $2 if $B$ guesses $H$, $M$ or $L$, respectively, no matter what the true value of the card is. Finally, another player $C$ has access to the value of the card and must guess what $B$ will guess. If his guess matches $B$'s, he wins $10, otherwise he gets nothing. Furthermore, $B$ can see $C$'s choice before making her decision. The MAID for this game can be seen in Figure 2.

This game can be solved using a MAID solution algorithm. The algorithm of [Koller and Milch 2001] uses a *relevance graph*. Each strongly connected component of the relevance graph is converted into an extensive form game tree. Unfortunately, in our game the relevance graph contains a single connected component, so the entire game must be converted into a tree. This tree must split at $A$'s, $B$'s and $C$'s choices, resulting in $3^3 = 27$ leaf nodes.

Can we do better? In our card game one might reason as follows: Agent $A$'s utility is not affected by the value of the card $J$, thus—according to WD strategies—$A$ will not condition his decision on the value of $J$ that he observes. In other words, $A$ is unmotivated and has no reason to tell $B$ the truth about $J$. Knowing that, $B$ will not believe $A$. Similarly $B$ will ignore what $C$ tells him, because—again—$C$'s utility is unaffected by the value of $J$ and thus he will ignore it in his decision. The new, simplified MAID can be seen in Figure 3.

Our algorithm in this instance would proceed as follows: During the first iteration, in the identification phase $A$ would be replaced by a chance node (none of the reasoning patterns hold for it) and the edge $(J, A)$ would be removed. In the pruning phase the edge $(J, C)$ is removed, because it is d-separated from $U_C$ given $C$. Furthermore, since $A$ is d-separated from $U_B$ given $\{B, C\}$ the edge $(A, B)$ would be removed as well. For the same reason $(C, B)$ would be pruned. During the second iteration no change is performed and the algorithm terminates. We now have to solve just two game trees, one for $B$ and one for $C$, with 9 leaf nodes each.

Now extend this simple game by adding more $C$-type players. Each of these would have access to $J$, win upon matching $B$'s choice and have his decision known to $B$. With $n$ such $C$-type players, the original game has a tree representation of $3^{(2+n)}$ leaves, whereas running the algorithm for one iteration would simplify this to $n+1$ trees of $3^2$ leaves each; savings in computing the equilibrium of the game are thus exponential.

## 5 ENUMERATING REASONING PATTERNS

We can easily tweak the algorithm to have it enumerate reasoning patterns as well. This can be done by *not* short-circuiting the **if** of line 10 of the algorithm and changing the functions **df**, **man**, **sig** and **rev** so as not to return when a reasoning pattern is found, but to add it to a list and iterate until no more reasoning patterns of that type are found.

As an example, consider the following two-stage principal-agent game with reputation. In each stage of the game the principal $P$ wants the agent $D$ to execute a task. The agent may be of type *good* or *bad*, but this type is not visible to the principal. The agent's decision is to exert *high* or *low* effort. Bad-type agents enjoy a greater utility from exerting low effort and vice versa for good-type agents. Moreover, the price offered by the principal also affects the agent's decision. Upon the completion (adequate or poor) of the task in the first stage, the principal moves on to offer another price to the agent for the second task. However, the agent's choice (visible performance) in the first stage is used to update a reputation value ($r_0$ to $r_1$), which the principal takes into consideration in his second offer. The reputation at each stage can be *high*, *medium* or *low* and is more likely to change between rounds when it is *medium*. The MAID for the game can be seen in Figure 4 (perfect recall edges are omitted for clarity). Running our algorithm reveals a set of interesting reasoning patterns (the "story" behind each reasoning pattern is provided by us and is *not* automatically generated by the workings of the algorithm):

- Direct effect: All nodes $P_1, P_2, D_1, D_2$ have direct effect. This corresponds to the fact that the players' actions have an immediate effect on their payoff.

- Manipulation (principal → agent): At each stage the principal manipulates the agent through the price offered. A higher price makes effort $= high$ more preferable for the agent and thus indirectly increases the principal's utility.

- Manipulation (agent → principal): $D_1$ manipulates $P_2$. This reflects the agent's thinking at stage 1: pretending to be *good* when in fact he is not can cause $P_2$ to offer him a high price for the second-round task (because the principal trusts him). In other words, this pattern corresponds to deceiving the principal and "milking" your reputation in the second round.

- Signaling (agent → principal): $D_1$ signals his *type* to $P_2$ by choosing an appropriate effort level. The variable *type* is not observed by the principal; however, the principal does care about it, since it provides him with insight to the agent's reasoning. The agent, on the other hand, wants his action to reveal something about this hidden piece of information. For example, if the agent is *good*, he wants the principal to know this so that he is offered a high price in the following round.

- Signaling ($r_0$ to $D_1$): $P_1$ signals the current reputation value to $D_1$. This is subtle but interesting: Suppose the agent believes that the principal has a current reputation value for him (the agent) that is *low*. In that case the agent might consider it a "lost cause" to exert *high* effort, since the principal on the next round is still going to believe he is *bad* and offer him a low price. Conversely, if the agent thinks his current $r_0$ is *high*, he has less reason to actually try and retain this reputation by exerting *high* effort. Thus, if $r_0$ is *medium*, the principal wants the agent to know it.

- Revealing-denying (*type* to $P_2$): $P_1$, by making a suitable offer to $D_1$ in the first round, can elicit information about his *type* that will be useful to $P_2$. For example, if the offer in the first round is too high, then it might be optimal for the agent to always exert *high* effort, which reveals little about his *type*. Experimenting with moderate offers allows the principal to risk guaranteed good performance in the first round, so that he may discover more about the agent's true *type* and exploit that in the second round.

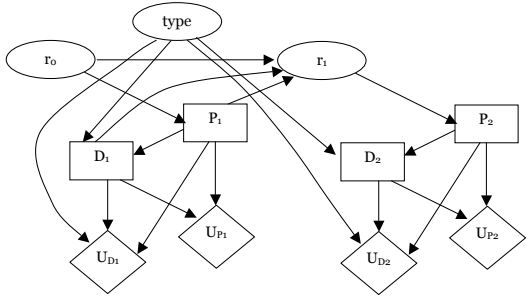

Figure 4: A two-stage principal-agent game

We see in this example how the reasoning patterns can be used to discover the possible motivations or patterns of thought agents may have when considering their decisions. Among them, there are some motivational patterns that are not obvious or easy to detect by means of any reasoning tool available today.

# 6 DISCUSSION

The algorithm presented in this paper uses the assumption of well-distinguishing (WD) strategies to simplify a game for the purpose of calculating a Nash equilibrium. Despite this being a reasonable assumption for computational agents, it is not clear whether people in general condition their actions only on observations that have an effect on their utilities, or whether they heed otherwise irrelevant signals. We plan to investigate this experimentally.

Another limitation of our algorithm is that it uses strictly graphical properties to discover and remove non-motivated nodes. Although this greatly simplifies the process and keeps it computationally efficient, a more rigorous examination of the parameters inside chance and utility nodes could lead to further simplifications. For example, it might be the case that a *signaling* reasoning pattern does hold graphically, but the parameters of the MAID are such that the first agent (Alice) always optimally decides on one particular action for all possible values of the signal, in which case the second agent (Bob) has no reason to infer anything about it by observing her action and he should rationally ignore it. Another example is zero-sum games, where selfish rational agents will optimally have no reason to signal any variable of interest to another agent, or believe any such signal.

Moreover, it has to be noted that our algorithm does not quantify the reasoning patterns it discovers. We cannot say, "According to this reasoning pattern, taking action $X$ will increase your utility by such and such an amount." It is up to the decision maker to determine the semantic interpretation and relative importance of the reasoning patterns.

Finally, although in the worst case exponential savings are possible, it is not yet clear how the algorithm performs in a *typical case*. We chose to avoid directly answering this question, mainly because of the challenge in defining "typical cases." Moreover, one should keep in mind that the algorithm's usefulness increases with the size of the game, since in large MAIDs deciding which observation is truly relevant to a decision is not obvious. In such cases, because the algorithm is polynomial—whereas Nash equilibrium computation is not, assuming PPAD $\neq$ P—, it can be run at negligible cost, even if the savings obtained are not significant.

## 7 CONCLUSION & FUTURE WORK

We have presented an algorithm for identifying reasoning patterns in games. This algorithm can be used in two ways: First, it can identify unmotivated decisions in a MAID graph; these can be effectively ignored for the purposes of calculating a Nash equilibrium, which can sometimes lead to considerable computational savings. Second, it is capable of discovering non-obvious patterns of thought agents might use when making their decisions. These patterns can be presented to a human decision maker to help him or her make good decisions. They may also used as inputs to other algorithms for decomposing, analyzing, explaining or even predicting agent or human behavior.

We believe that in complex systems, and especially those in which human and computer agents interact heavily, it is important for successful agents to model the "context" of their interaction. Certain features of games are likely to bring about different motivations in human behavior, such as reciprocity or cooperation. Reasoning patterns are promising as modeling aids for three reasons: First, they are adequately rich to analyze arbitrarily complex games, yet they are concise in their number. Second, they have a behavioral flavor, in that they talk about *reasoning* and not just utility maximization, yet they also have a rigorous mathematical and normative foundation. And third, they are discoverable in efficient (polynomial) ways. We plan to extend this line of research with the ultimate goal of constructing computerized agents who have a better understanding of human motivations and are better aligned with their goals. This will entail calculating semantic characterizations of the various reasoning patterns, looking at their graphical properties as well as the parameters within the nodes, and explicitly calculating the implications of each pattern to the agents' optimal or equilibrium strategies.

## Acknowledgments

The research reported in this paper was supported in part by AFOSR MURI grant FA9550-05-1-0321.